\documentclass[12pt]{iopart}
\usepackage{graphicx}

\begin{document}
\title{Weak localization and transport gap in graphene antidot lattices}
\author{J Eroms, D Weiss}
\address{Universit\"at Regensburg, Institut f\"ur Experimentelle und Angewandte Physik, D-93040 Regensburg, Germany}
\ead{jonathan.eroms@physik.uni-regensburg.de}
\submitto{\NJP}
\begin{abstract}
We fabricated and measured antidot lattices in single layer graphene with lattice periods down to 90~nm. In large-period lattices, a well-defined quantum Hall effect is observed. Going to smaller antidot spacings the quantum Hall effect gradually disappears, following a geometric size effect. Lattices with narrow constrictions between the antidots behave as networks of nanoribbons, showing a high-resistance state and a transport gap of a few mV around the Dirac point. We observe pronounced weak localization in the magnetoresistance, indicating strong intervalley scattering at the antidot edges. The area of phase-coherent paths is bounded by the unit cell size at low temperatures, so each unit cell of the lattice acts as a ballistic cavity.
\end{abstract}

Since its discovery in 2004~\cite{Bib:Novoselov2004}, graphene and its peculiar relativistic dispersion relation have inspired experimentalists and theorists alike~\cite{Bib:GeimRise}. In the past few years, many fundamental effects have been demonstrated in graphene, often with specific modifications due to the properties of Dirac fermions. For example, the quantum Hall effect in single layer graphene shows steps at values $(n+1/2)\,4e^2/h$ (where $n$ is an integer), which is caused by the chiral motion of the carriers~\cite{Bib:NovoselovQHE,Bib:KimQHE}. Weak localization was found to be suppressed in some samples~\cite{Bib:Morozov}, and is generally more complex than in diffusive metals, since the strong isospin-orbit coupling leads to a destructive interference of time-reversed paths~\cite{Bib:McCann,Bib:Tikhonenko}. The interplay of several scattering mechanisms then results in very rich physics, from weak localization to antilocalization, or complete suppression of both. Electronic confinement in nanoribbons turns out to be linear in the inverse constriction width, much stronger than in materials with a quadratic dispersion~\cite{Bib:HanGNR}.

In our study, we turn to a periodic nanoscale structure, namely antidot lattices~\cite{Bib:DWAntidots}. Here, a periodic lattice of holes is etched into a two-dimensional layer. In high-mobility two-dimensional electron gases, pronounced oscillations appear in the magnetoresistance, which is a consequence of the the electron motion in the chaotic antidot billiard~\cite{Bib:FGK}. With the mobilities available in graphene layers on oxidized silicon substrates, clear ballistic effects of this kind are not yet feasible. However, since the phase coherence length in graphene can approach $1\ \mu$m at $^4$He temperatures, we expect to see modifications of weak localization. Furthermore,  Pedersen {\em et al.} recently suggested to use graphene-based antidot lattices to create a transport gap and confine electrons in a tailor-made point defect in such a lattice~\cite{Bib:PedersenADs}. As we will show, our antidot lattices show clear signatures of electron paths confined to phase-coherent unit cells of the lattice and exhibit a high-resistance state around the Dirac point.

\begin{figure}
\begin{center}\includegraphics[width=10cm]{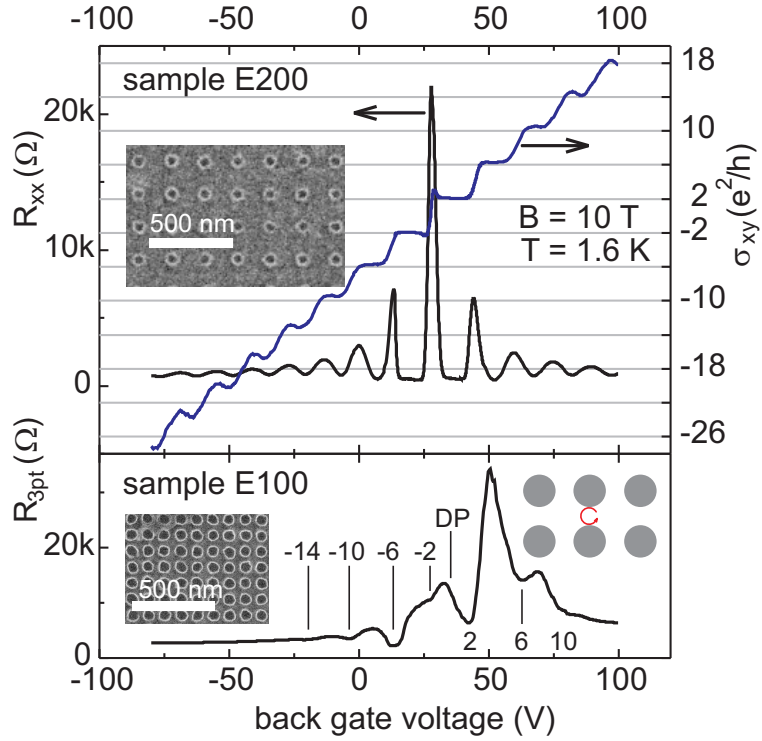}\end{center}
\caption{Quantum Hall effect in single layer graphene antidot lattices. Upper panel: Lattice period $a=200$~nm. A large number of oscillations is visible. Lower panel: $a=100$~nm, 3-terminal resistance. Here, the quantum Hall effect can only be observed if a complete cyclotron orbit can pass between the antidots (sketch in lower right inset). The scanning electron micrographs show the antidot pattern in samples E200 and E100. DP marks the position of the Dirac point.}
\label{Fig:QHE}
\end{figure}

\begin{table}\caption{\label{Tab:Samples}Overview of the antidot samples and their geometry. Length and width refer to the patterned region of the flake.}
\begin{indented}
\item[]\begin{tabular}{@{}lllll}
\br
Sample name&lattice period $a$&antidot diameter $d$&length&width\\
\mr
A140&140 nm&\phantom{1}65 nm&2.2 $\mu$m &1.3 $\mu$m \\ 
B140&140 nm&\phantom{1}80 nm&3.4 $\mu$m &2.5 $\mu$m \\ 
B200&200 nm&\phantom{1}70 nm&3.2 $\mu$m &2.5 $\mu$m \\
C200&200 nm&150 nm&5.0 $\mu$m &1.7 $\mu$m \\
D90&\phantom{1}90 nm&\phantom{1}54 nm&5.1 $\mu$m &2.9 $\mu$m \\ 
D100&100 nm&\phantom{1}60 nm&5.1 $\mu$m &2.9 $\mu$m \\
D110&110 nm&\phantom{1}65 nm&5.0 $\mu$m &2.9 $\mu$m \\
E100&100 nm&\phantom{1}62 nm&4.0 $\mu$m &4.0 $\mu$m \\ 
E200&200 nm&\phantom{1}67 nm&4.0 $\mu$m &4.0 $\mu$m \\
E400&400 nm&\phantom{1}64 nm&4.0 $\mu$m &4.0 $\mu$m \\
\br
\end{tabular}
\end{indented}
\end{table}

\begin{figure}
\begin{center}\includegraphics[width=10cm]{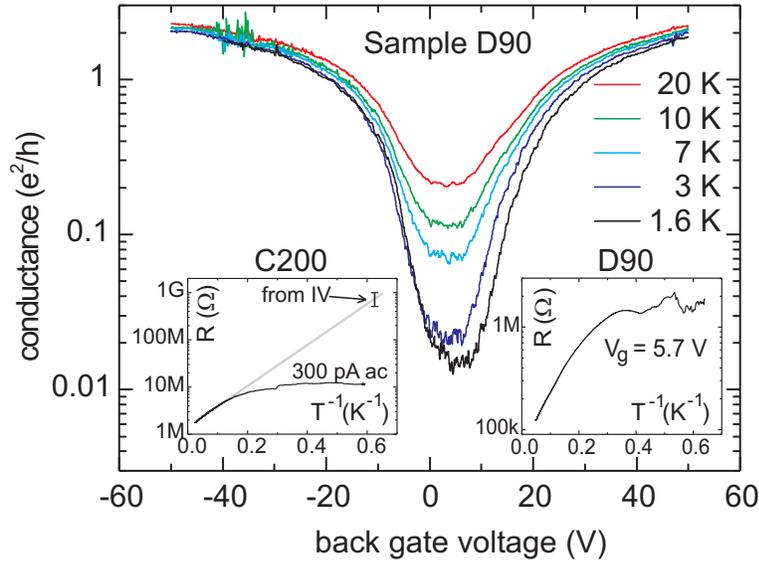}\end{center}
\caption{Insulating behaviour of sample D90 (lattice period $a=90$~nm) at various temperatures. The conductance at 1.6~K can be varied by 2 orders of magnitude in the accessible gate voltage range. Insets: the temperature dependence of samples C200 (with very large antidots) and D90 at the Dirac point measured with an ac bias of 300~pA and 1~nA, respectively. The zero bias data point of C200 at 1.6~K, taken from an $I$-$V$-curve, shows that the thermally activated behaviour continues to low temperatures. The thin gray line corresponds to thermal activation with a 22~K energy gap.}
\label{Fig:Gap}
\end{figure}

\begin{figure}
\begin{center}\includegraphics{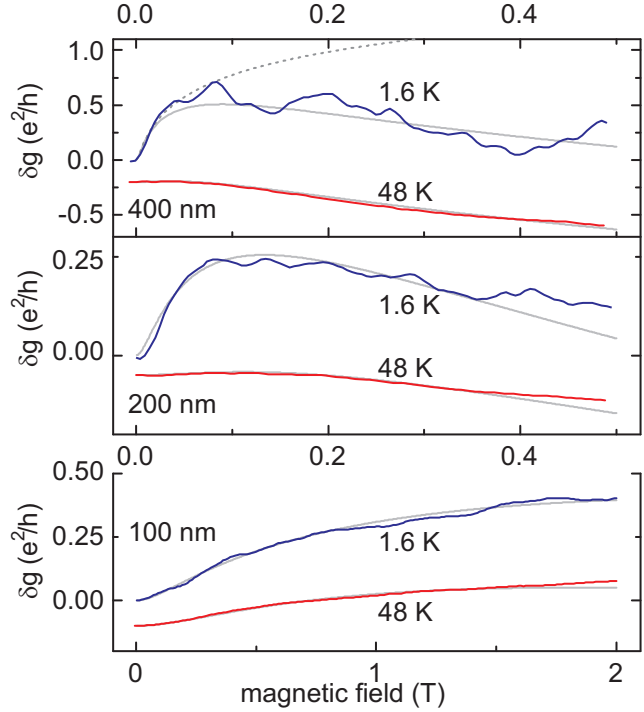}\end{center}
\caption{Weak localization in samples E400, E200, and E100. Note the different field scale for sample E100. The thin gray line is a fit to equation~\ref{Eq:McCannWL}, the dashed line is a fit to the WL correction in a dirty semiconductor, with the same phase coherence length. The graphs at 48~K were shifted vertically for clarity.}
\label{Fig:WLperiods}
\end{figure}

\begin{figure}
\begin{center}\includegraphics{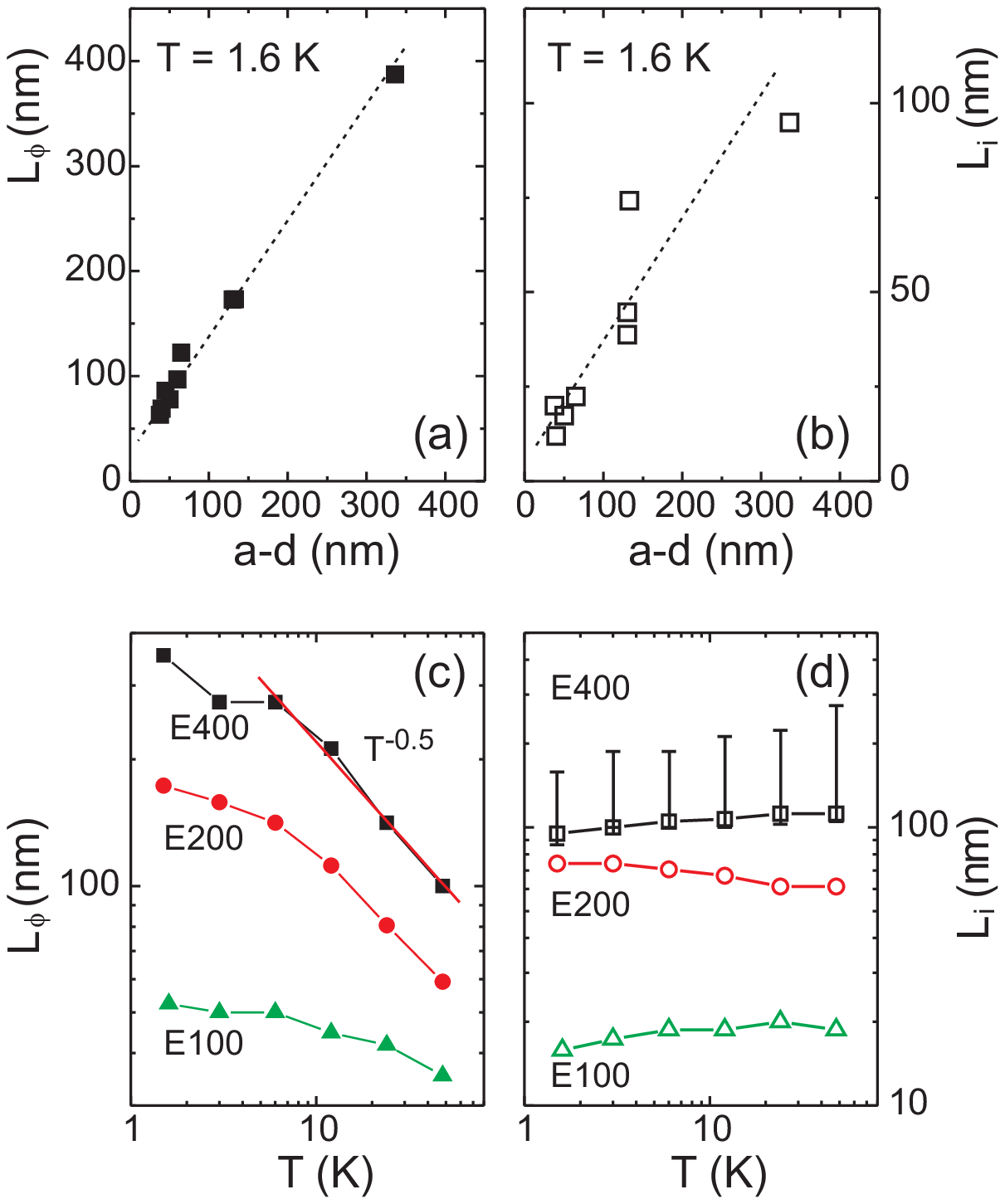}\end{center}
\caption{(a) and (b): Geometry dependence of the phase coherence length (a) and the intervalley scattering length (b) from the WL fit for all samples in this study. The distance $a-d$ refers to the width of the constrictions between the antidots. The dashed lines are guides to the eye. (c) and (d): Temperature dependence of $L_\phi$ (c) and $L_\mathrm{i}$ (d) of sample E100, E200, and E400. The error bars on the data of E400 in subfigure (d) give the range of $L_\mathrm{i}$ when $L_*$ is varied between 70 nm and 700 nm.}
\label{Fig:Tdep}
\end{figure}

The graphene flakes were prepared by exfoliating natural graphite with the adhesive tape technique described in \cite{Bib:Novoselov2004}. A highly doped Si-wafer with 300 nm of thermally grown silicon dioxide was used as a substrate. Single layer flakes were identified by the contrast of a grayscale image in the optical microscope~\cite{Bib:Blake}. We verified the reliability of this method with Raman spectroscopy and quantum Hall measurements. In total, five single layer flakes (A to E) were selected. We patterned contacts with electron beam lithography using pre-defined markers, metal evaporation and lift-off. The metalization was either Ti/AuPd (flakes A to D) or Pd (flake E). The antidot lattices were prepared with O$_2$ reactive ion etching (flakes A to D) or Ar ion beam etching at 350~V (flake E). The sample names consist of the flake and the antidot lattice period $a$ in nm. Table 1 lists the lattice period $a$ and the antidot diameter $d$ of all the samples in this study. The length and width of the patterned region are also given.

The measurements were performed in a $^4$He-cryostat at temperatures between 1.6~K and 48~K using lock-in techniques with an ac bias current of 10~nA or lower. The carrier density was varied using the Si wafer as a back gate. Baking the sample {\em in situ} at 150$^\mathrm{o}$C for several hours improved the homogeneity of the samples and moved the Dirac point closer to zero back gate voltage.

Figure~\ref{Fig:QHE} shows a quantum Hall measurement at $B=10$~T on flake E. Sample E200 (and also E400, not shown here) shows clear Hall plateaus at quantized values $(n+\frac{1}{2})\frac{4e^2}{h}$ and Shubnikov-de Haas oscillations in the longitudinal resistance, as expected for single layer graphene. While in E200 a large number of oscillations is visible, sample E100\footnote{During that measurement, only 3 contacts were working, so we could only measure the 3-terminal resistance. It always includes one contact resistance and either $R_{xx}$ or $R_\mathrm{2pt}$, depending on the magnetic field direction and carrier polarity~\cite{Bib:Datta}. Here, at gate voltages below the Dirac point, the signal is $R_{xx}$-like and we observe SdH-oscillations. The poor visibility of the $\nu=-2$ feature is due to an inhomogeneity, which also appears at $B=0$ at the same back gate voltage.} only shows minima at filling factors $\nu =2$, 6, 10, and 14, i.e., at low carrier densities. Calculating the diameter of a cyclotron orbit $2R_\mathrm{C} = 2\hbar k_F/eB$ at the last visible filling factor, $\nu = 14$ yields $2R_\mathrm{C} = 40$~nm, which corresponds to the lithographic width $a-d$ of the constrictions between the antidots. At higher densities, the cyclotron orbits are larger and cannot move freely along the sample, therefore the quantum Hall effect is suppressed. In sample D90, with even smaller constrictions, only the plateau at $\nu=2$ is visible (not shown). The mobility in our flakes was ranging between $\mu=5000$~cm$^2$/Vs and $\mu=7000$~cm$^2$/Vs. For example, at a carrier density of $10^{12}$~cm$^{-2}$ this leads to a mean free path between 60~nm and 80~nm. Therefore, the mean free path was not sufficient to allow the observation of clear commensurability effects in the antidot lattices. Shen {\em et al.} recently claimed the observation of commensurability effects in graphene antidot lattices on SiC~\cite{Bib:Peide}, although without an independent determination of the carrier density.

Now we turn to the behaviour around the Dirac point. In large-area graphene, the conductivity at the Dirac point does not drop to zero, but stays at a finite value. A graphene field-effect transistor can therefore never be turned off. In graphene nanoribbons (GNR), lateral confinement and edge disorder lead to a high-resistance state \cite{Bib:HanGNR,Bib:Mucciolo}, allowing to operate a GNR transistor even at room temperature \cite{Bib:RTGNR}. The on-state current in those devices is, of course, limited by the nanoribbon width. An antidot lattice can be considered as a network of many nanoribbons in parallel, and hence the maximum current can be increased by upscaling the device. Our antidot lattices also show a high-resistance state around the Dirac point. Figure~\ref{Fig:Gap} shows the conductance of sample D90 at various temperatures, measured with a lock in technique using an ac bias current of 0.7~nA and zero dc bias. The on/off ratio at 1.6~K is about 100.  Since the constrictions in our devices are of the order of a few tens of nm, the observed gap is rather small. From $I$-$V$-measurements on sample C200, we extract a gap of around 6~mV along the whole antidot lattice\footnote{We cannot determine if the voltage drop is distributed evenly between the antidot lattice cells, or if the voltage drops predominantly at the narrowest constriction.}. The resistance at the Dirac point is thermally activated with an activation gap of about 22~K$\cdot k_\mathrm{B}$ for samples C200 and D90. Since the lock-in measurement was performed with a finite ac current bias, the resistance apparently saturates once the sample voltage exceeds the size of the gap. Taking the true zero bias resistance from the full $I$-$V$-curve, we see that in fact the thermal activation continues down to 1.6~K (see left inset in figure~\ref{Fig:Gap}).

All our samples show very clear weak localization (WL) in the magnetoresistance (see figure~\ref{Fig:WLperiods}). This is in strong contrast to the situation in unpatterned graphene flakes of similar size, where conductance fluctuations often mask the localization feature completely and averaging over a range of gate voltages is necessary to unveil WL~\cite{Bib:Tikhonenko,Bib:Kechedzhi}. More importantly, due to the strong isospin-orbit coupling in graphene the WL contribution is suppressed. WL only appears if the rate of intervalley scattering $\tau_\mathrm{i}^{-1}$(which breaks the isospin-orbit coupling) is higher than the phase-breaking rate $\tau_{\phi}^{-1}$. Finally, the scattering rate $\tau_*^{-1} = \tau_\mathrm{i}^{-1} + \tau_z^{-1} + \tau_w^{-1}$ (where $\tau_w^{-1}$ is due to trigonal warping and $\tau_z^{-1}$ is the intravalley scattering rate) also enters into the description of WL in graphene~\cite{Bib:McCann}.
The conductivity correction $\delta g$ in diffusive, bulk graphene is governed by the following equation:
\begin{equation}\label{Eq:McCannWL}
    \delta g=\frac{e^2}{\pi h} \left(F\left(\frac{B}{B_\phi}\right) - F\left(\frac{B}{B_\phi+2B_\mathrm{i}}\right) -2F\left(\frac{B}{B_\phi+B_*}\right)  \right)
\end{equation}
where $F(z)=\ln z + \Psi(0.5 + z^{-1})$, $B_{\phi,\mathrm{i},*} = \hbar\tau_{\phi,\mathrm{i},*}^{-1}/4eD $ and $\Psi$ is the digamma function~\cite{Bib:McCann}. $D$ is the diffusion constant, and $\tau_{\phi,\mathrm{i},*}$ are the respective scattering times. The corresponding lengths are $L_{\phi,\mathrm{i},*} = \sqrt{D\tau_{\phi,\mathrm{i},*}}$.
The first term in equation~\ref{Eq:McCannWL} leads to a positive magnetoconductance and is exactly compensated by the second term, unless $B_\mathrm{i}$ is large, i.e., strong intervalley scattering is present. The first term is also identical to the magnetic field dependence of the WL in a dirty, two-dimensional semiconductor in the limit where the mean free path is much shorter than the phase coherence length~\cite{Bib:BeenakkerVanHouten}. In principle, fitting the experimental WL curves to equation~\ref{Eq:McCannWL} allows the determination of three scattering rates by looking at the behaviour over a large field range. However, other contributions to magnetoresistance, e.g., slight inhomogeneities or quantum effects at large fields can mask the WL correction and make the fit less reliable. In practice, therefore only the phase breaking rate can be determined with high accuracy since it is the only parameter leading to an upturn of the magnetoconductance. In our case, a satisfactory fit can be obtained with the scattering length $L_*$ varying between 70 nm and 700 nm. The intervalley scattering length $L_\mathrm{i}$ then simultaneously varies by about a factor of 1.7, where a shorter $L_*$ leads to a longer $L_\mathrm{i}$ or vice versa. The phase-coherence length $L_\phi$, however, is always unambiguously defined, regardless of the values chosen for $L_*$ and $L_\mathrm{i}$, since it is taken from the positive magnetoconductance around $B=0$.

Figure~\ref{Fig:WLperiods} shows the conductivity correction $\delta g$ (obtained by subtracting the zero-field conductance from the magnetoconductance)  of the samples E400, E200, and E100 at 1.6~K and 48~K and a voltage of 20~V applied to the back gate. Contrary to unpatterned graphene samples of mesoscopic size~\cite{Bib:Tikhonenko}, in the antidot lattices the WL feature is clearly visible without averaging over a range of gate voltages. We fitted the data using equation~\ref{Eq:McCannWL}, and also, for comparison, to the expression for a dirty semiconductor. Both fits describe the positive magnetoconductance at low fields, giving the same phase coherence length. Obviously, the theory of a diffusive semiconductor fails to reproduce the graphene data at higher fields, while the fit to equation~\ref{Eq:McCannWL} is very good over the entire field range.

Let us first consider intervalley scattering.
Since a WL feature is visible, intervalley scattering must be present in our samples. The etched boundaries of the antidots are an obvious source of intervalley scattering. This can be seen by comparing the data for different lattice periods at 1.6~K and 48~K and looking at the temperature dependence of $L_\phi$ (see figure~\ref{Fig:Tdep}). For samples E400 and E200, $L_\phi$ decreases as $T^{-0.5}$ at high temperatures. At 48~K, most electron paths will not reach an antidot edge within a phase-coherence length. Intervalley scattering is therefore not effective to make the WL correction visible. Indeed, the experimental data of E400 only shows a very faint WL feature at 48~K, in E200, WL at 48~K is visible but weak. Sample E100, where $L_\phi$ allows to explore the entire unit cell even at 48~K, correspondingly shows a well defined WL feature at 48~K.

In figures~\ref{Fig:Tdep}~b and \ref{Fig:Tdep}~d we show the geometry and temperature dependence of $L_\mathrm{i}$ as fitting parameter in equation~\ref{Eq:McCannWL}. The error bars in the data for sample E400 in figure~\ref{Fig:Tdep}~d show the interdependence of $L_\mathrm{i}$ and $L_*$. For example, the data point at $T=1.6$~K extends from 90~nm to 160~nm. The lower value for $L_\mathrm{i}$ is obtained by taking $L_*$ between 200~nm and 700~nm, whereas $L_\mathrm{i}=160$~nm is found by taking $L_*=70$~nm. Comparing $L_\mathrm{i}$ and the constriction width $a-d$  (which is the shortest length scale imposed by patterning) for all samples, we find that the intervalley scattering length $L_\mathrm{i}$ from the diffusive fit is always much shorter than $a-d$. For example, for sample E400 with a lattice period of 400~nm the diffusive theory yields $L_\mathrm{i}\approx 100$~nm. We do not expect that additional sources of intervalley scattering, other than the antidot edges, were introduced during fabrication, since the graphene in between the antidots was protected by about 100~nm of PMMA resist during etching. Also, flake D contained a reference section without antidots and showed a much longer intervalley scattering length of 550~nm (see below). Instead, we believe that the short $L_\mathrm{i}$ is an artefact of the diffusive theory, which assumes that the mean free path is much shorter than any other length scale of the system. In our samples, the mean free path is between 60~nm and 80~nm, and the antidots, as sources of intervalley scattering, are introduced on a similar length scale. In the diffusive theory, to get a clear WL feature, $L_\mathrm{i}$ must be much shorter than $L_\phi$, whereas in our situation each charge carrier hitting an antidot is scattered between the valleys and WL is restored. A recent numerical study supports this view: In a tight-binding simulation a ballistic cavity with either abrupt scattering at the edges (strong intervalley scattering) or with mass confinement (no intervalley scattering)
was considered. The magnetoconductance only shows WL when intervalley scattering occurs at the cavity boundary~\cite{Bib:Wurm}. The WL correction to the magnetoconductance in a phase-coherent chaotic cavity is given by a Lorentzian~\cite{Bib:Baranger}, whose width corresponds to the flux through the cavity area. Within the experimental uncertainty, the measured data can be fitted both with the ballistic and the diffusive theory. Fitting a Lorentzian (i.e., the result of the ballistic theory) with the diffusive theory gives a phase coherence length of the order of the square root of the cavity area but the intervalley scattering length is much shorter than any of the cavity dimensions. We therefore conclude that the poor correspondence of the values for $L_\mathrm{i}$ with any of the relevant lengths in our samples shows that our samples are better described by the ballistic theory than the diffusive theory.

Now we turn to the phase coherence length. In a truly ballistic sample, the WL correction should follow a Lorentzian~\cite{Bib:Baranger} rather than the shape described  by equation~\ref{Eq:McCannWL}. However, since equation~\ref{Eq:McCannWL} contains 3 parameters, it is next to impossible to distinguish between both theories within the experimental uncertainty. Taking $L_\phi=\sqrt{A}$, where $A$ is the characteristic area from the Lorentzian fit, both theories give the same results when fitting the experimental WL correction. In an unpatterned reference section on flake D, using equation~\ref{Eq:McCannWL}, we determined $L_\phi=700$~nm and $L_\mathrm{i}=550$~nm at 1.6~K, i.e., larger than all the lattice periods we fabricated. In the antidot lattices, the phase coherence length is limited by the lattice geometry as can be seen in figure~\ref{Fig:Tdep}~a. More precisely, at 1.6~K, the lattice period sets a limit to the phase coherence length, while in samples with a large antidot spacing the phase coherence length falls off as $T^{-0.5}$ at higher temperatures. Since the WL correction actually probes an area distribution of time-reversed paths we believe that the electron paths are confined to the ballistic unit cells of the antidot lattice, which are only weakly coupled by the nanoribbon-like constrictions. Alternatively, one might argue that the phase coherence is simply cut off by scattering at the antidot edges. If that were the case, no closed coherent paths could be formed, since the electrons would lose their phase memory when hitting an antidot, and WL would not be observed. The fact that we do observe a pronounced WL feature shows that electrons experience intervalley scattering at the antidot edges, which is essential to observing WL in graphene, but retain their phase, otherwise the closed time-reversed paths could not interfere. Therefore we believe that at low temperatures and small lattice spacings, our samples are best described as a network of loosely coupled ballistic cells, where the enclosed area of phase-coherent paths is bounded by the cavity area.

In summary, our antidot lattices show rich behaviour going from almost unperturbed graphene with a well-defined quantum Hall effect to a two-dimensional network of loosely coupled, phase-coherent cavities. As we go to smaller lattice spacings, the quantum Hall effect gradually disappears while a high-resistance state forms around the Dirac point. The pronounced WL feature shows that the antidot edges introduce intervalley scattering while conserving phase coherence of closed, time-reversed paths.

\ack
We would like to thank M. Hirmer, T. Korn and C. Sch\"{u}ller for the Raman measurements, and \.{I}. Adagideli, C. Beenakker, J. Bundesmann, D. Horsell, E. Mucciolo, K. Richter, M. Wimmer and J. Wurm for fruitful discussions.

\clearpage


\begin{thebibliography}*

\bibitem{Bib:Novoselov2004}  Novoselov K S,  Geim A K,  Morozov S V, Jiang D,
Zhang Y, Dubonos S V, Grigorieva I V and Firsov A A 2004 {\em Science} {\bf 306} 666

\bibitem{Bib:GeimRise}  Geim A K and Novoselov K S 2007 {\em Nature Materials} {\bf 6} 183

\bibitem{Bib:NovoselovQHE} Novoselov K S, Geim A K, Morozov S V, Jiang D, Katsnelson M I, Grigorieva I V, Dubonos S V and A. A. Firsov 2005 {\em Nature} {\bf 438} 197

\bibitem{Bib:KimQHE} Zhang Y, Tan Y-W, Stormer H L and Kim P 2005 {\em Nature} {\bf 438} 201

\bibitem{Bib:Morozov} Morozov S V, Novoselov K S, Katsnelson M I, Schedin F, Ponomarenko L A, Jiang D and Geim A K 2006 {\em Phys. Rev. Lett.} {\bf 97} 016801

\bibitem{Bib:McCann} McCann E, Kechedzhi K, Fal'ko V I, Suzuura H, Ando T and Altshuler B L 2006 {\em Phys. Rev. Lett.} {\bf 97} 146805

\bibitem{Bib:Tikhonenko} Tikhonenko F V, Horsell D W , Gorbachev R V and Savchenko A K 2008 {\em Phys. Rev. Lett.} {\bf 100} 056802

\bibitem{Bib:HanGNR} Han M Y, \"{O}zyilmaz B, Zhang Y and Kim P 2007 {\em Phys. Rev. Lett.} {\bf 98} 206805

\bibitem{Bib:DWAntidots}Weiss D, Roukes M L, Menschig A, Grambow P, von Klitzing K and Weimann G 1991 {\em Phys. Rev. Lett.} {\bf 66} 2790

\bibitem{Bib:FGK} Fleischmann R, Geisel T and Ketzmerick R 1992 {\em Phys. Rev. Lett.} {\bf 68} 1367

\bibitem{Bib:PedersenADs} Pedersen T G, Flindt C, Pedersen J, Mortensen N A, Jauho A-P and Pedersen K 2008 {\em Phys. Rev. Lett.} {\bf 100} 136804

\bibitem{Bib:Blake} Blake P, Hill E W, Castro Neto A H, Novoselov K S, Jiang D, Yang R, Booth T J and Geim A K 2007 {\em Appl. Phys. Lett.}  {\bf 91} 063124


\bibitem{Bib:Datta}Datta S 1995 {\em Electronic Transport in Mesoscopic Systems}, (Cambridge: Cambridge University Press)

\bibitem{Bib:Peide} Shen T, Wu Y Q, Capano M A, Rokhinson L P, Engel L W and Ye P D 2008 {\em Appl. Phys. Lett.}  {\bf 93} 122102

\bibitem{Bib:Mucciolo} Mucciolo E R, Castro Neto A H and Lewenkopf C H 2009 {\em Phys. Rev. B}  {\bf 79} 075407

\bibitem{Bib:RTGNR} Wang X, Ouyang Y, Li X, Wang H, Guo J and Dai H 2008 {\em Phys. Rev. Lett.} {\bf 100} 206803

\bibitem{Bib:Kechedzhi} Kechedzhi K, Horsell D W, Tikhonenko F V, Savchenko A K, Gorbachev R V, Lerner I V, and Fal'ko V I 2009 {\em Phys. Rev. Lett.} {\bf 102} 066801

\bibitem{Bib:BeenakkerVanHouten} Beenakker C W J and van Houten H, 1991, {\em Solid State Physics} {\bf 44} 1

\bibitem{Bib:Wurm} Wurm J, Rycerz A, Adagideli \.{I}, Wimmer M, Richter K and Baranger H U 2009 {\em Phys. Rev. Lett.} {\bf 102} 056806

\bibitem{Bib:Baranger} Baranger H U, Jalabert R A and Stone A D 1993 {\em Phys. Rev. Lett.} {\bf 70} 3876







\end{thebibliography}
\end{document}